\documentclass[sigconf]{acmart}
\usepackage{graphicx}
\usepackage{xcolor}
\usepackage{comment}

\usepackage{booktabs}   %% For formal tables:
\usepackage{subcaption} %% For complex figures with subfigures/subcaptionso
\usepackage{listings}
\usepackage{framed}
\usepackage{url}
\usepackage{mdwlist} 
\usepackage{enumitem}

\setlist{nolistsep}
\lstdefinestyle{myCStyle}{
  language=C,
  basicstyle=\tiny,
  showspaces=false,
  showstringspaces=false,
  identifierstyle=\color{blue},
  commentstyle=\color{green!60!black},
  stringstyle=\color{orange},
  keywordstyle=\color{purple!80!black}
}

\setcopyright{rightsretained}
\acmDOI{}

\acmISBN{}

\acmConference[]{}{}{}
\acmYear{}
\copyrightyear{}

\begin{document}

\title{Refactoring the MPS/University of Chicago Radiative MHD (MURaM) Model for GPU/CPU Performance Portability Using OpenACC Directives}
%\titlenote{}

%% \subtitle{Extended Abstract}
%% \subtitlenote{The full version of the author's guide is available as
%%   \texttt{acmart.pdf} document}

%\author{}
%\authornote{Dr.~Trovato insisted his name be first.}
%\affiliation{%
 % \institution{University of Delaware}
 % \streetaddress{101 Smith Hall}
  %\city{Newark}
  %\state{Delaware}
  %\postcode{19701}
%}
%\email{}

%\author{}
%\authornote{Dr.~Trovato insisted his name be first.}
%\affiliation{%
 % \institution{Max Planck Institute for Solar System Research}
  %\streetaddress{Justus-von-Liebig-Weg 3}
  %\city{Gottingen}
  %\state{Germany}
  %\postcode{37077}
%}
%\email{}

%\author{}
%\authornote{Dr.~Trovato insisted his name be first.}
%\affiliation{%
%  \institution{NCAR}
%  \streetaddress{P.O. Box 2008}
%  \city{Boulder}
 % \state{Colorado}
 % \postcode{80305}
%}
%\email{}

\author{Eric Wright}
  \affiliation{
    \institution{University of Delaware}
    \streetaddress{}
    \city{Newark}
    \state{Delaware}
    \postcode{}
  }
  \email{efwright@udel.edu}
  
\author{Damien Przybylski}
  \affiliation{
    \institution{Max Planck Institute for Solar System Research}
    \streetaddress{}
    \city{Gottingen}
    \state{Germany}
    \postcode{}
  }
  \email{przybylski@mps.mpg.de}  
  
\author{Matthias Rempel}
  \affiliation{
    \institution{National Center of Atmospheric Research}
    \streetaddress{}
    \city{Boulder}
    \state{Colorado}
    \postcode{}
  }
  \email{rempel@ucar.edu}
  
\author{Cena Miller}
  \affiliation{
    \institution{National Center of Atmospheric Research}
    \streetaddress{}
    \city{Boulder}
    \state{Colorado}
    \postcode{}
  }
  \email{cmille73@ucar.edu}
  
\author{Supreeth Suresh}
  \affiliation{
  \institution{National Center of Atmospheric Research}
    \streetaddress{}
    \city{Boulder}
    \state{Colorado}
    \postcode{}
  }
  \email{ssuresh@ucar.edu}
  
\author{Shiquan Su}
  \affiliation{
    \institution{National Center of Atmospheric Research}
    \streetaddress{}
    \city{Boulder}
    \state{Colorado}
    \postcode{}
  }
  \email{shiquan@ucar.edu}
  
\author{Richard Loft}
  \affiliation{
    \institution{National Center of Atmospheric Research}
    \streetaddress{}
    \city{Boulder}
    \state{Colorado}
    \postcode{}
  }
  \email{loft@ucar.edu}

\author{Sunita Chandrasekaran}
  \affiliation{
    \institution{University of Delaware}
    \streetaddress{}
    \city{Newark}
    \state{Delaware}
    \postcode{}
  }
  \email{schandra@udel.edu}

  %Eric, Damien, Matthias, Cena, Supreeth, Shiquan, Rich, Sunita

% \renewcommand{\shortauthors}{R. Searles et al.}

\begin{abstract}
The MURaM (Max Planck University of Chicago Radiative MHD) code is a solar atmosphere radiative MHD model that has been broadly applied to solar phenomena ranging from quiet to active sun, including eruptive events such as flares and coronal mass ejections. 
The treatment of physics is sufficiently realistic to allow for the synthesis of emission from visible light to extreme UV and X-rays, which is critical for a detailed comparison with available and future multi-wavelength observations. 
This component relies critically on the radiation transport solver (RTS) of MURaM; the most computationally intensive component of the code. 
The benefits of accelerating RTS are multiple fold: A faster RTS allows for the regular use of the more expensive multi-band radiation transport needed for comparison with observations, and this will pave the way for the acceleration of ongoing improvements in RTS that are critical for simulations of the solar chromosphere.
We present challenges and strategies to accelerate a multi-physics, multi-band MURaM using a directive-based programming model, OpenACC in order to maintain a single source code across CPUs and GPUs. 
Results for a $288^3$ test problem show that MURaM with the optimized RTS routine achieves 1.73x speedup using a single NVIDIA V100 GPU over a fully subscribed 40-core Intel Skylake CPU node and with respect to the number of simulation points (in millions) per second, a single NVIDIA V100 GPU is equivalent to 69 Skylake cores.
We also measure parallel performance on up to 96 GPUs and present weak and strong scaling results. 

\end{abstract}

\maketitle

%% LaTeX sources

\section{Overview}
The MURaM (Max Planck University of Chicago Radiative MHD) code \citep{Vogler2005simulations,Rempel:2014:SSD,Rempel:2017:corona} is one of the primary solar models used for simulations of the upper convection zone, photosphere (visible surface of the sun) and corona. 
MURaM simulations have contributed substantially to our understanding of solar phenomena ranging from the origins of quiet sun magnetism, the structure and evolution of sunspots and active regions, to solar flares and the initiation of coronal mass ejections. MURaM also plays a key role in interpreting high resolution solar observations. With the construction of the Daniel K. Inouye Solar Telescope (DKIST), an NSF investment~\cite{dkist} exceeding \$300 million, resolution of ground based observational solar physics is poised to take an order of magnitude leap forward. % \textcolor{DFP - This sentence does not really fit, since the chromosphere has not even been meantioned yet. Delete it? or move it down to the chromosphere section?: These new high resolution observations will allow us to observe the chromosphere in greater detail than ever before. }

 Radiation transport is a key component in realistic models of the solar atmosphere. In order to capture the frequency dependence of opacity in a cost effective way, spectral lines are combined in bands according to their strength \citet{Nordlund:1982} (and sometimes also frequency) and average opacities are computed for each band (typically 4-12). The radiation transport is then performed for each band instead of hundred thousands of individual frequency points. But even with this substantial simplification, radiation transport in multiple bands can consume more than 90\% of the computing time for photospheric simulations. As a consequence many simulations have been performed with a single "gray" band, which limits direct comparability with observations. A speedup of radiation transport will make multi-band radiation transport possible as the routine option. However, radiation transport becomes even more challenging when applied in the chromosphere.
%Originally based on a magnetohydrodynamics (MHD) module from the University of Chicago, MURaM is jointly developed and used by the National Center for Atmospheric Research (NCAR), the Max Planck Institute for Solar System Research (MPS) and the Lockheed Martin Solar and Astrophysics Laboratory (LMSAL).
 %MURaM simulations have contributed substantially to our understanding of solar phenomena ranging from the origins of quiet sun magnetism, the structure and evolution of sunspots and active regions, to solar flares and the initiation of coronal mass ejections. MURaM also plays a key role in interpreting high resolution solar observations. With the construction of the Daniel K. Inouye Solar Telescope (DKIST), an NSF investment (http://dkist.nso.edu) exceeding \$300 million, resolution of ground based observational solar physics is poised to take an order of magnitude leap forward. 
%These new high resolution observations will allow us to observe the chromosphere in greater detail than ever before.

The solar chromosphere, lying between the photosphere and the transition to the corona, is one of the least understood parts of the Sun. It is difficult to observe; as it is only visible in the cores of strong lines where the signal to noise ratio is low and the magnetic field diagnostics available are relatively poor. High resolution observations with DKIST will soon allow us to observe the chromosphere in greater detail than ever before. The theoretical treatment of the chromosphere is equally challenging, as none of the usual assumptions can be used. Neither radiation, nor the ionisation state of atoms can be treated in local thermodynamic equilibrium (LTE), in which the source function is simply given by the temperature dependent Planck function and population levels can be calculated with the Saha-Boltzman equation~\cite{aguilera2007multi}. 

The primary requirement to correctly model the structure and dynamics of the chromosphere is the inclusion of non-equilibrium atomic populations, primarily hydrogen. Because the timescales of hydrogen ionisation and recombination in the chromosphere are of the same order as the dynamical timescales, it is a non-equilibrium problem. This requires evolving the ionisation state of the atoms as they are advected by bulk plasma motions, and then calculation of the excitation and de-excitation by collisional and radiative processes. A module to track the non-equilibrium evolution of hydrogen, using a simplified 1D radiation field, is now available for the MURaM code and increases the computational cost by about a factor of 5. Extending the module to perform realistic simulations of the chromosphere will require a thorough treatment of the radiation field, including multiple wavelengths covering all the important atomic transitions.

%The primary requirement to correctly model the structure and dynamics of the chromosphere is the inclusion of non-equilibrium atomic populations, primarily hydrogen. Because the timescales of hydrogen ionisation and recombination in the chromosphere are of the same order as the dynamical timescales, it is a non-equilibrium problem. This requires tracking the ionisation state of the atoms through the advection, and excitation and de-excitation by collisional and radiative processes. The ionisation of helium becomes important in the energy balance of the upper chromosphere. Additionally, the non-equilibrium behaviour of helium, and other elements such as calcium and magnesium, may be important to understand observations of chromospheric spectral lines. A module to track the non-equilibrium evolution of hydrogen is now included in the MURaM code and increases the computational cost by a factor of four. Extending this to include Helium or other elements will require a more thorough treatment of the radiation field, increasing the computational costs by up to an order of magnitude. \textcolor{red}{Damien: "....more thorough treatment of the radiation field..." - does this signify the multi-spectral band concept? If yes it would be good to introduce that here}

\begin{figure}[h]
  \centering
      \includegraphics[width=2.5in]{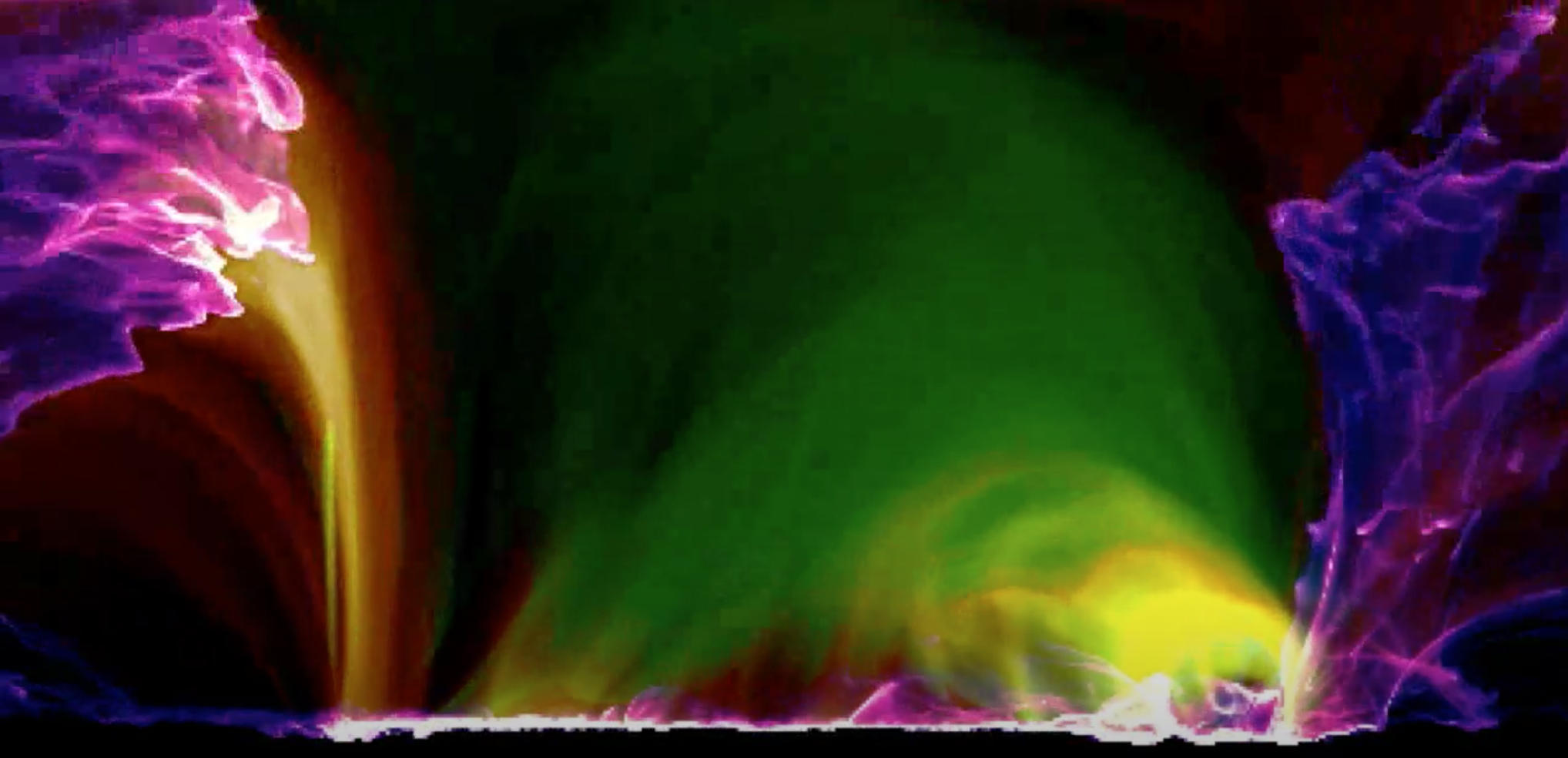}
  \caption{Data inspired simulation of a solar eruption. Presented is a still image from this outreach movie~\cite{movie} that is based on  \citet{Cheung:etal:2019:Flare}.}
    \label{fig:solarflare}
\end{figure}

The problem under study, MURaM is capable of simulating the coupled solar atmosphere from the upper convection zone into the lower solar corona, covering a density stratification of more than 25 scale heights. This makes MURaM a crucial tool for studying how magnetic field emerging from the solar interior is energizing the solar atmosphere and is leading to rapid release of energy in form of flares and coronal mass ejections. Figure~\ref{fig:solarflare} shows an example of such a simulation from \citet{Cheung:etal:2019:Flare}. Currently these simulations are "stand-alone" setups that are inspired by solar events, but do not aim at modeling observed solar events in detail. Using these simulations in the future as a utility to study the solar drivers of space weather events will require data ingestion through boundary driving or full data assimilation and the ability to run such simulations in real-time. 

In summary, future science applications of the MURaM code will focus on detailed studies of the coupled solar atmosphere as well as data-driven simulations of solar events. This requires the combination of (1) higher resolution, (2) more sophisticated physics in terms of radiation transport and (3) ensemble simulations. The resulting increased demand in computing speed cannot be achieved with traditional CPU based systems.
\vspace{-1.2em}
\subsection{Motivation}
%\textcolor{blue}{
%This section will talk about the motivations of why we want to port MURaM to GPUs. We will discuss the science goals, the need for more computation power, any other solar physics or related codes that are running on GPU, and why we are using OpenACC specifically (over OpenMP, and instead of CUDA).
%}
The above scientific advancements can be achieved by a combination of two classes of simulations: (1) High resolution simulations in smaller domains with more detailed physics that allow for the direct comparison with observations that will be available from DKIST. Here it is specifically critical to {\bf speed up radiation transport} such that multi-band simulations become the norm and more detailed physics can be added in the near future. (2) Low resolution simulations in large domains with the standard physics of MURaM in order to simulate (and predict) observed solar space weather events. For these simulations it is critical to {\bf reach the real-time threshold} in order to allow for data-assimilative simulations in the future. Starting from the current CPU baseline of the code, an increase of computational capabilities by 1-2 orders of magnitude is needed. 

To achieve these ambitious science goals, migration from petascale to exascale computing is clearly required. In the context of the US Department of Energy (DoE)-led exascale program~\cite{exascale}, this means focusing on Graphics Processing Units (GPUs). There are several possible language/compiler pathways to exascale computing with GPUs: vendor-specific languages like CUDA~\cite{CUDA}, directive-based standards such as OpenMP offloading model above~\cite{openmp} and OpenACC~\cite{openacc}; and domain-specific languages (DSL) such as Stella and its successor GridTools~\cite{gysi2015stella}. 

The project to refactor MURaM for GPUs had multiple goals. First and foremost, we sought to improve the throughput of the model, as measured by site updates per second, or simulated seconds per wall clock second to the maximum extent possible.  The project's target throughput speedup was parity between one NVIDIA V100 GPU and 100 Intel Xeon v4 processor cores~\cite{searles2018abstractions}. Experience refactoring similar models suggested that this speed-up was achievable for well-implemented code with sufficient data parallelism to saturate the GPUs. Our secondary goal was maintainability, by preserving portability of the model between CPUs and GPUs to the maximum extent possible. Finally, we sought, as a tertiary goal,  performance-portability: namely, finding a maintainable implementation that did not compromise (significantly) model performance on CPUs or GPUs.  At the time this project began in early 2018, the directive-based parallelization approach  appeared to offer the best prospect of meeting the first two goals, while performance portability was an open question. Of the directive-based systems,  OpenACC~\cite{openacc} was deemed to be the most production ready in 2018. 

%SEEMS LIKE A NON SEQUITUR: 
%Applying an accelerated MURaM model of the coupled photosphere/corona system to solar eruptive events will be instrumental in improving the inner boundary condition of heliospheric simulations of space weather. 

%The MURAM code base is compact, containing up to 60,000 source lines of code (SLOC). The standard configuration run of MURAM has only 10,000 SLOC. Hence our first goal is to provide a GPU-enabled version for the standard configuration. To that end we need to prepare the code such that they are ready to tap into what GPUs can offer.
 
%\begin{figure*}
%    \centering
%    \includegraphics[width=0.96\textwidth]{HMI_AIA_300000.png}
%    \caption{Highlighting MURaM capabilities; Example of a MURaM flux emergence simulation - \textbf{Top Row:} left panel: synthetic photospheric magnetogram, middle panel: the vertical velocity, right panel: magnetic field at the coronal base. 
%    \textbf{Middle Row:} Synthetic AIA emission in the 131, 193, and 94 pass bands from a top view computed from MURaM. \textbf{Bottom Row:} Same as middle row but from a side view. The snapshot corresponds to the time right after a flare, AIA 131 and 94 show hot post flare loops, 193 the footpoints of these loops in the transition region (flare ribbons).}
%    \label{fig1}
%\end{figure*}

\vspace{-1em}

The directive-based programming model, OpenACC has been widely used in the past several years to migrate large scale applications demonstrating maturity and stability with the compiler implementations of the high level features. Some of the large (including production) applications that uses OpenACC include MPAS~\cite{yang2019accelerating}, ANSYS~\cite{ansys}, Icosahedral non-hydrostatic (ICON)~\cite{sawyer2014towards}, LQCD Monte Carlo~\cite{bonati2018portable}, and VASP~\cite{maintz2018strategies}. % and other miniapplications such as minisweep~\cite{searles2019mpi,vergara2020experiences}

OpenMP~\cite{chapman2008using} also a directive-based programming model was predominantly targeting shared memory architectures since its inception in 1999. Since 2013, OpenMP has been supporting accelerators. While some applications such as GenASiS~\cite{budiardja2019targeting}, Pseudo-Spectral Direct Numerical Simulation-Combined Compact Difference (PSDNS-CCD3D)~\cite{yeung} QMCPack~\cite{kim2018qmcpack}, GAMESS~\cite{kwack2019performance}, and other mini-application such as minisweep~\cite{vergara2020experiences}, SU3~\cite{davis2020performance} have been ported to the OpenMP offloading model, we observed that the compiler implementations of the offloading features~\cite{diaz2019analysis} are not yet fully mature and stable enough to tackle MURaM especially when we started the code development of MURaM in early 2018. To that end, MURaM uses OpenACC~\cite{chandrasekaran2017openacc}.

\subsection{Contributions}
\begin{itemize}[leftmargin=*]

    \item Specific GPU algorithmic enhancements, namely: asynchronous programming, loop fusion, and array replication, to the method of discrete ordinates used in MURaM's radiation transport model. Radiation transport is the single most expensive part of most astrophysical codes, so these optimizations are broadly applicable across astrophysics. 

    \item Acceleration of the radiation transport in MURaM will in turn make routine use of multi-band radiation transport possible in solar physics models. 
    
    \item Multi-band radiation transport will advance understanding of the solar chromosphere when combined with non-equilibrium treatment of atomic populations.
    
    \item Demonstration of the efficacy of the use of OpenACC directive-based approach to achieve performance-portability across CPUs and GPUs in a solar physics model. 
    
\end{itemize}

%\section{Motivation}

\section{MURaM Routines}
\label{sec:routines}
The primary focus of the MURaM code, a solar physics code developed over the past 2 decades, are the detailed studies of the solar atmosphere with sufficient realism to allow for forward modeling of synthetic observables from visible to EUV and X-ray emission for direct comparison with a wide range of solar observations. To this end the MURaM code is typically applied to small Cartesian local regions on the Sun with lateral and vertical simulation domains ranging from a few $1000$~km to more than $100,000$~km, with typical numerical grid spacings in a range of $2-200$~km (for comparison, the solar radius is about $700,000$~km).  
The code combines a fourth order conservative MHD scheme with short characteristics radiation transport as described in~\cite{vogler2005simulations}. 
The MURaM MHD scheme uses a cell centered finite difference approach, the $\nabla\cdot\vec{B}=0$ constraint is enforced through hyperbolic divergence cleaning \citep{Dedner:etal:2002:divB}. 
%The MHD scheme is stabilized through the use of artificial diffusivities, that are based on a slope-limited diffusion scheme as described in \citet{Rempel:2014:SSD}. The code uses a tabulated non-ideal equation of state for a solar element mixture \citep[e.g. OPAL,][]{Rogers:opal:1996}. 
The radiation transport resolves the angular dependence of the radiation field by computing rays in (typically) 24 directions based on a Carlson quadrature \citep{Carlson:1963}. 
%the frequency dependence of the opacity is considered through a multigroup approach that follows the opacity binning approach introduced by \citet{Nordlund:1982} and adapted for use in MURaM by \citet{Voegler:etal:2004a}. 
Radiation transport is an intrinsically non-local problem, which poses implementation challenges on distributed memory architectures. The short characteristics solver of MURaM treats radiation transport locally, by using intensities from a previous time step (or iteration) as the starting point for the ray integration on each shared memory block. Achieving global convergence requires typically 3 to 4 iterations in the radiation transport solve.  
The above formulation has been extensively used to study magneto-convection in the solar photosphere and upper convection zone. The code has been expanded to also include the overlying solar corona~\cite{rempel2009radiative}. 
These implementations lead to a fully explicit code that can be parallelized for shared memory systems through domain decomposition and MPI communication.
The major computational routines of the MURaM code, there algorithms and purpose, are summarised below:
\begin{itemize}[leftmargin=*]
    \item {\bf MHDRES} - Calculate the right hand side of the MHD equations in conservative form. Calculate the derivatives of the fluxes using a fourth order central difference scheme.
    \item {\bf TVDDIFF} - Numerical diffusion required to stabilise the solution. Calculated using a slope limited diffusion scheme.
    \item {\bf EOS} - Using the density and energy from the MHD solution calculate the equation of state variables; temperature, pressure, electron number, entropy.  
    \item {\bf RTS} - Calculate the radiation field using the thermodynamic variables of the current MHD snapshot. A short-characteristics algorithm is used. Consists of 4 main functions:
    \begin{itemize}[leftmargin=*]
        \item Interpolate - Perform trilinear interpolation of the MHD variables from the MHD grid to the staggered RTS grid.  Calculate the radiation source function and opacities to be used in the integration routines.
        \item Integrate - Integrate along each ray of the quadrature.
        \item  Exchange - Communicate intensity information to downstream processors in the ray direction.
        \item Flux - From the specific intensity calculate the average intensity and radiative fluxes.
    \end{itemize}
    \item {\bf DIVBCLEAN} -  Diffuse and disperse numerical $\vec{\nabla}\cdot\vec{B}$ errors through a hyperbolic $\vec{\nabla}\cdot\vec{B}$ cleaning approach.
    \item {\bf INTEGRATE} - Calculate the updated variables for the next stage of the Runge Kutta algorithm using the divergence of fluxes from {\bf MHDRES} and additional source terms such as gravity radiative heating/cooling.
    \item {\bf DST} - Exchange subdomain ghost cells with neighbours.
    \item {\bf SYNC} - Determine the maximum time-step and synchronise the timestep between subdomains.
    \item {\bf VLIM} - Dynamically adjust the velocity, energy and Alfv{\'e}n velocity limits to prevent extreme cells causing overly restrictive timestep constraints.
     \item {\bf BOUNDARY} - Implement the vertical boundary conditions; A stratified open boundary at the bottom with passive field advection. Upper boundary is open to outflows, the magnetic field is matched to a potential magnetic field. The latter requires fast Fourier transforms.
\end{itemize}

\section{Profiling}
Code profiling reveals several metrics that show in-depth information about code performance and behaviors. There are various profiling applications that serve a variety of purposes. We have used GNU gprof, ARM MAP and NVIDIA nvprof tools during the entirety of this development process. Since the beginning of this project nvprof has been deprecated by NVIDIA and replaced with NVIDIA NSight Compute, which offers many of the same functionalities. For this paper version since many of our experiments were conducted before NSight Compute's release, we discuss our results with nvprof.
%\vspace{-1.2em}

\subsection{CPU Profiling}
\label{sec:cpu_profiling}
To effectively port and optimize the performance of a code it is important to gain a high-level view of the code's performance as-is to identify areas that are most computationally intensive and consume the greatest percentage of the total runtime. Additionally, observing the scalability of the CPU code will give insight into any overheads caused by MPI communication.

Two key metrics observed during CPU profiling would be the percentage of time spent in each routine considering a single MPI rank on a single CPU core and the MPI communication of using many CPU cores as well as multiple compute nodes. Using ARM MAP for the single core profile enabled the creation of a graphical function call graph that showed the flow of code execution as well as the percentage of runtime taken by each step. Nvprof also allows for some CPU profiling, however was primarily used to observe CPU performance side-by-side with GPU performance. ARM MAP is also used to capture the MPI communication when using multiple MPI ranks across several CPUs.

\begin{figure}[t]
\centering
    \includegraphics[width=1.00\columnwidth]{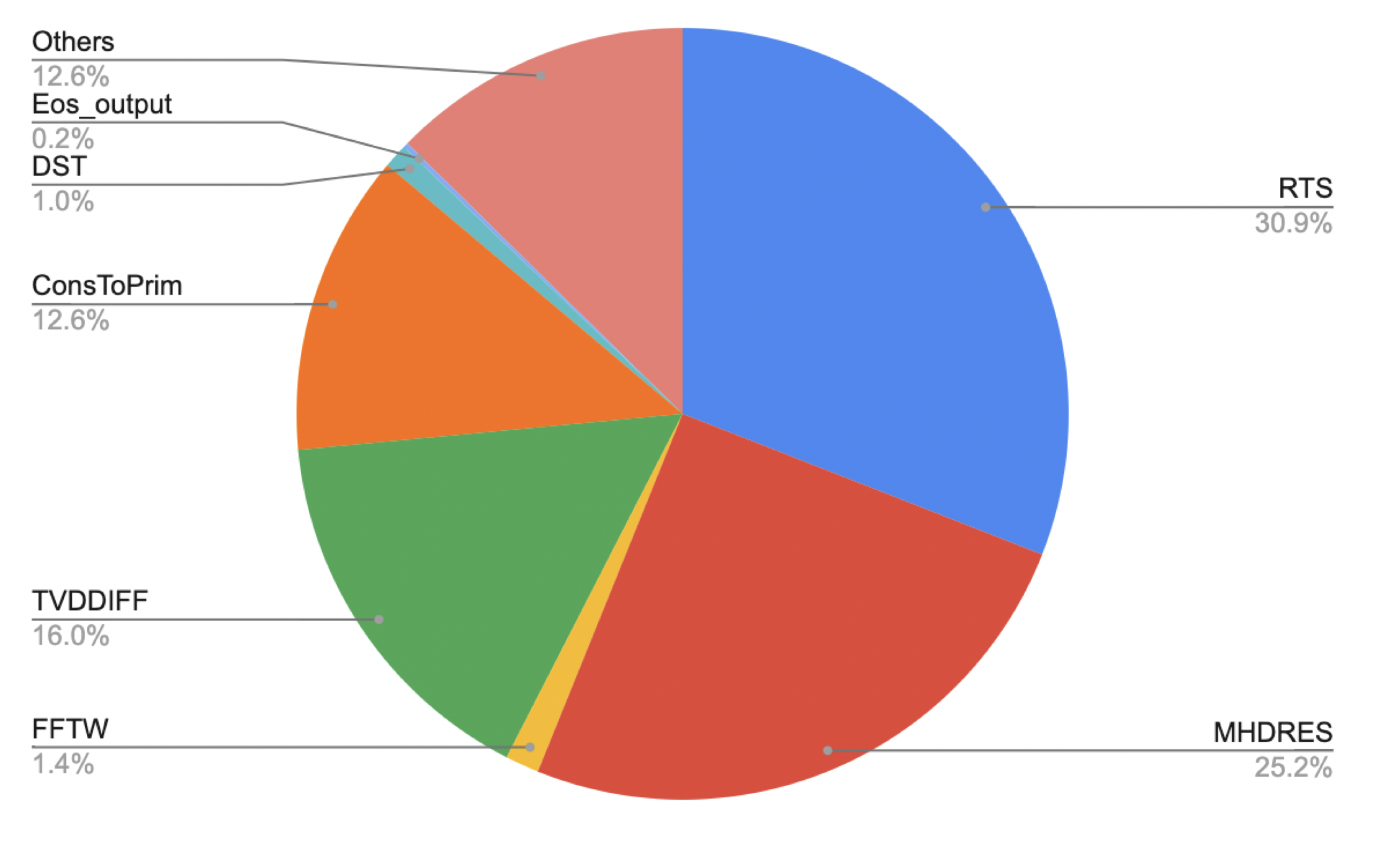}
    \caption{Single core profiling.}
    \vspace{-1.5em}
    \label{fig:single core profiling}
\end{figure}

The single core profiling results are shown in Figure ~\ref{fig:single core profiling}. From these results, there are three portions of the code that account for ~71\% of the total runtime: \textit{MHDRES}, \textit{TVDDIFF} and \textit{RTS}. These profiling results are for single-band radiation transport. In multi-band applications RTS could be up to 12 times more expensive. The importance and functionality of these routines are described in Section ~\ref{sec:routines}. Since this is only showing performance of a single CPU core it gives insight as to what routines are computationally intensive without concerns of any MPI overhead. With these results we have a clear starting point of what routines should be immediately targeted for parallelization.

\textit{RTS} is significantly more complicated that \textit{MHDRES} and \textit{TVDDIFF} as the radiation portion of the code contains many smaller subroutines. The general techniques used to parallelize \textit{MHDRES} and \textit{TVDDIFF}, as well as the majority of the rest of the code, is described in detail in Section ~\ref{sec:implementation}. The parallelization of \textit{RTS} presented many interesting challenges and was handled very differently than the rest of the code. The parallelization of \textit{RTS} is described in Section ~\ref{sec:RTS}.

\begin{figure}[t]
\centering
    \includegraphics[width=1.00\columnwidth]{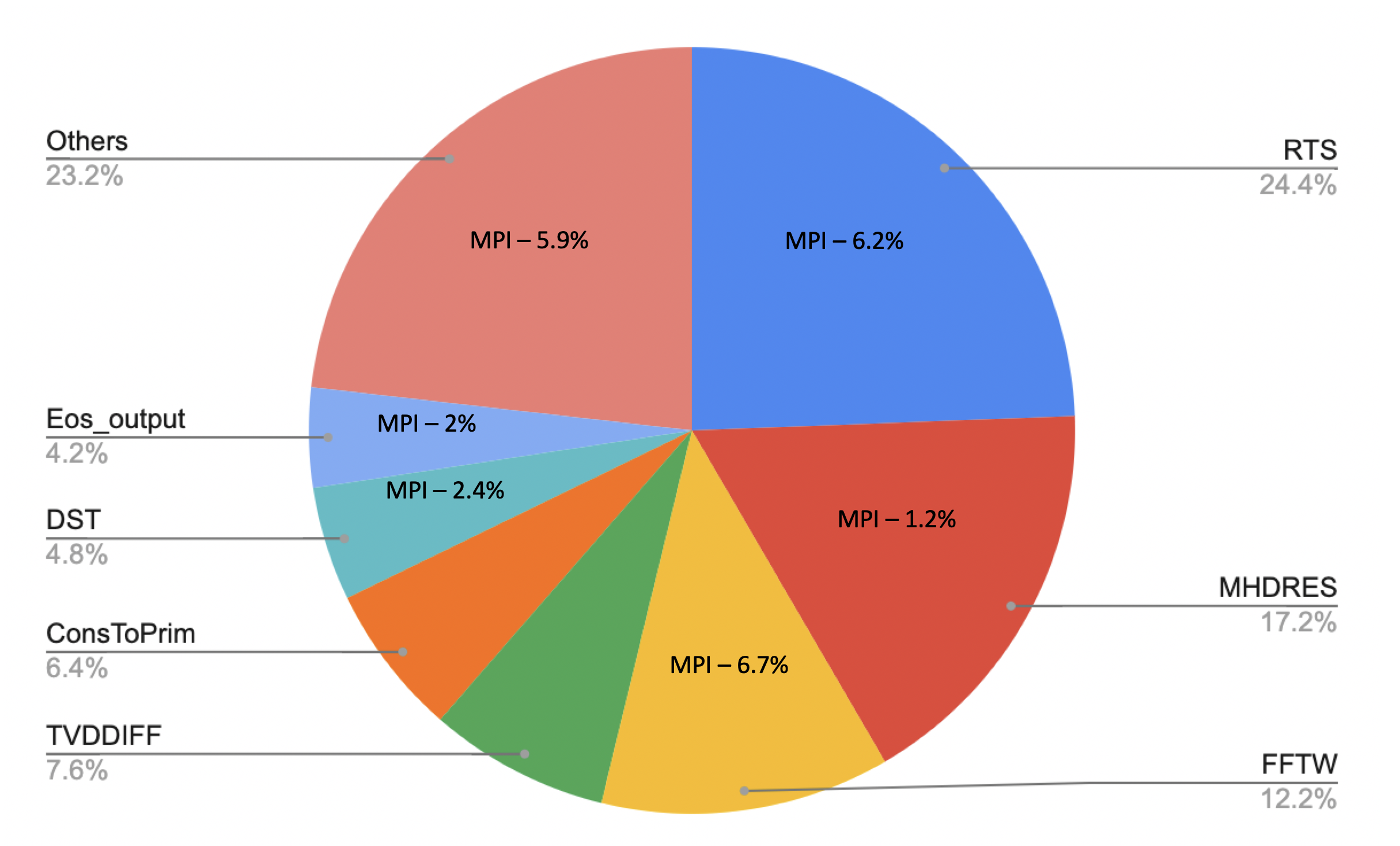}
    \caption{Multi core profiling.}
    \vspace{-1.5em}
    \label{fig:Multi core profiling}
\end{figure}

The MPI profiling using ARM MAP results are shown in Figure ~\ref{fig:Multi core profiling}. In contrast to the single core profiling, additional routines like \textit{FFTW} and \textit{DST} affects the total runtime, with \textit{FFTW} being the MPI version of the FFTW 3.8 library and \textit{DST} communicating ghost cells through MPI. \textit{RTS} also includes a significant amount of MPI communication. Figure ~\ref{fig:Multi core profiling} shows that the MPI communication within \textit{RTS} alone accounts for 6.2\% of the total runtime. Optimizing the MPI communication is an important task for the performance of both and CPU and GPU code. The inclusion of GPU-aware MPI for device-to-device direct MPI data transfers is discussed in Section ~\ref{sec:gpu_mpi} and some MPI optimizations specific to \textit{RTS} are discussed in Section ~\ref{sec:rts_variations}.

\subsection{GPU Occupancy}
\label{sec:gpu_occupancy}
GPU profiling tools can reveal important information about the achievable performance of a code. One performance metric that significantly guided our development process was the achieved and theoretical GPU occupancy. GPU occupancy for NVIDIA GPUs refers to the percentage of warps active at a given time. Theoretical GPU occupancy can be determined by examining the GPU register and shared memory usage when a kernel is compiled. Low theoretical GPU occupancy is generally caused by a kernel needing to use too many registers or too much shared memory, which results in the GPU not having enough resources to support using every warp simultaneously.

\begin{figure}[ht!]
\centering
    \includegraphics[width=0.8\columnwidth]{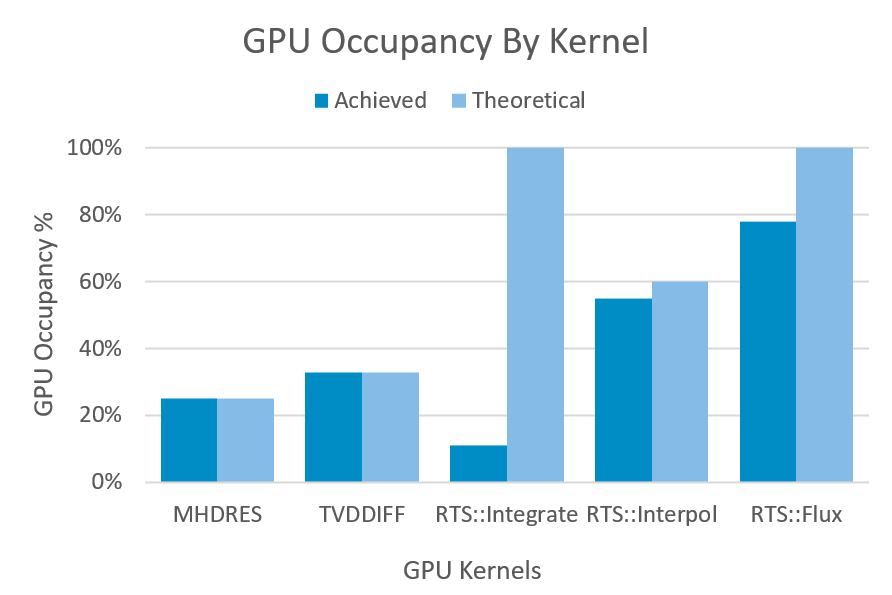}
    \caption{Theoretical and achieved GPU occupancy of various GPU kernels within MURaM.}
    \vspace{-1.5em}
    \label{fig:gpu_occupancy}
\end{figure}

Achieved GPU occupancy is simply the GPU occupancy that is observed during kernel execution. It is ideal to have theoretical GPU occupancy as close to 100\% as possible and to have achieved GPU occupancy as close to the theoretical GPU occupancy as possible. There are many possible reasons for achieved occupancy to be lower than theoretical occupancy, some of which was observed in our work with MURaM. The GPU occupancy of several important kernels is shown in Figure ~\ref{fig:gpu_occupancy} and using this metric we can identify two key performance issues within MURaM.

\begin{figure}[t]
\centering
    \includegraphics[width=0.85\columnwidth]{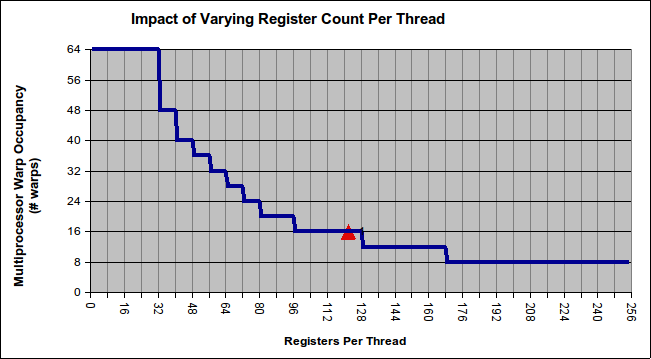}
    \caption{Effect of GPU register usage on GPU occupancy in the MHDRES and TVDDIFF kernels.}
    \vspace{-1.5em}
    \label{fig:register_occupancy}
\end{figure}

Firstly, many of the kernels are reaching very low theoretical occupancy, with \textit{MHDRES} and \textit{TVDDIFF} reaching 25\% and 33\% respectively. For these two kernels, the low theoretical occupancy is due to an over-allocation of GPU registers per thread. Figure ~\ref{fig:register_occupancy} shows the potential theoretical occupancy of the \textit{MHDRES} kernel when the registers per thread changes. If any more than 32 registers are allocated per thread the theoretical occupancy will go below 100\%, and since \textit{MHDRES} is compiled to use 122 registers per thread, it can only reach 25\% theoretical occupancy.

This is an interesting problem in directive-based programming models, such as OpenACC, because the programmer relies on the compiler to assign GPU registers when generating the GPU code. When using a device-specific language, such as CUDA, the programmer can fine-tune this register allocation to a higher degree. In the case of \textit{MHDRES} the PGI compiler has determined that 122 registers is the optimal number to achieve the most performance, and incorporating a hard register limit of 32 registers sees a significant performance decrease.

Secondly, \textit{RTS::integrate} shows a different problem where the theoretical occupancy is very high while the achieved occupancy is very low. This means that when \textit{RTS::integrate} is compiled the number of registers and the amount of shared memory assigned is such that the GPU could potentially support every warp running simultaneously. However, in practice only 10\% of those warps are used when executing the kernel.
This is due to a lack of parallelism being exposed in a way that we are only able to parallelize across two dimensions of our three dimensional domain. Typically, modern GPUs are expected to perform SIMT parallelism on several millions of data points, but in this kernel we are only hitting a few ten thousands, resulting in the 10\% GPU occupancy. The reasoning for this will be thoroughly elaborated on in Section ~\ref{sec:RTS}.
\vspace{-0.5em}

\section{Developing MURaM using OpenACC}
\label{sec:implementation}
%\textcolor{blue}{
%This section will cover our general techniques for parallelizing the code. We will talk about how we apply OpenACC directives, how we optimize data movement between kernels, how we validate correctness with the validation suite and PGI PCAST, and how we implemented and debugged CUDA-aware MPI. We can also talk about RTS parallelization in this section, but skip the parts related to the alternate versions of the integrate routine.
%}
In this section we provide an overview of our development process with OpenACC.

\subsection{Refactoring the MURaM Code}
%{\color{purple} RDL - Renamed (MURaM already was parallel - using MPI!), reordered and revised 4.1, since it jumps right into the refactoring process to the part where we start messing with directives. I've tried to make it clear that you first talk to the scientists, establish how to tell your code is scientifically correct and then start tampering. } 

Maintaining correctness during the migration of MURaM's large and complex code-base to support GPUs presented a challenge. To address this key issue, we chose an incremental, test-driven development approach throughout.  This involved three steps: 1) identifying suitable baseline test cases; 2) creating a correctness validation build-test system; and 3) incremental migration to GPUs by applying OpenACC directives and validation of the results. It is worth noting that steps 1 and 2 required close collaboration with the solar physicists on our team.
%Managing a large codebase while porting to GPUs can be fairly difficult, so we utilized a common development process for directive-based code porting. 

%{\color{purple} RDL - I think we benefit here by having Matthias or Cena write a short paragraph about how we selected test case, i.e. resource cost, functionality coverage and what our reference test case actually is - if that isn't somewhere else in this paper!}

%{\color{teal} CEM - (Matthias, I'll need help for specific science reasons it was chosen/parameter settings) 
%\textcolor{red}{SC: Rich, Updated text on testcase selection}
We chose a test case with a grid size of 192x64x64 for our validation suite because it's small enough to run on a single CPU core while still capturing the solar atmosphere from the upper convection zone into the lower solar corona and therefore testing all implemented physics in the code. We used this setup to generate the CPU reference data required to validate the GPU implementation against.
During the porting and development process, its small size allowed us to quickly and repeatedly run the model for the 50 time-steps required by the validation suite without exhausting our limited cluster resources. It’s also large enough to decompose and test in different x, y, and z core layouts. 

\begin{figure}[ht!]
\centering
    \includegraphics[width=0.5\columnwidth]{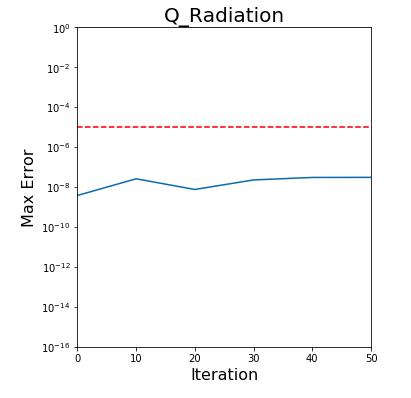}
    \caption{Maximum Relative Error in GPU calculated Q\_Radiation over 50 iterations.}
    \vspace{-1.5em}
    \label{fig:Q_Rad_ErrorvsIter}
\end{figure}

\begin{figure*}[ht]
\centering
    \includegraphics[width=1.3\columnwidth]{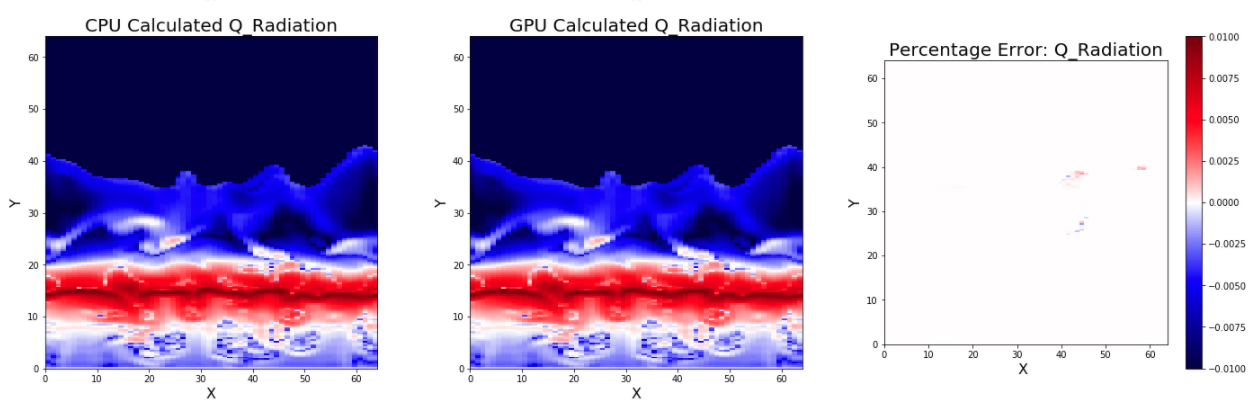}
    \caption{Percentage Error in GPU calculated Q\_Radiation after 50 iterations.}
    \vspace{-1.5em}
    \label{fig:CPU_GPU_PercError}
\end{figure*}

To validate correctness, we developed a validation suite which builds and runs the ported code using different combinations of processor layouts. At specific timesteps, diagnostic variable data is output into files. The test diagnostics can be compared to reference diagnostics generated by the CPU master code using a matching data set and processor layout. We defined an acceptance tolerance as the variance observed between MPI CPU runs with varying decompositions and core layouts, which is a maximum relative error margin of 1e-05. However, bit errors on the order of 1e-7 can cause large relative errors at points where the reference data is close to zero. To handle this issue, an absolute error between the diagnostic and reference data is also calculated and accepted when less than 1e-7.  Figures ~\ref{fig:Q_Rad_ErrorvsIter} and ~\ref{fig:CPU_GPU_PercError} show some of the graphs generated by our validation suite. These figures are often very helpful in debugging problems that occur at predictable areas of the domain, such as at processor boundaries.

For additional correctness checking we utilized a new feature in the PGI compiler called PGI Compiler Assisted Software Testing (PCAST)~\cite{pcast}. This allows direct comparison of data from a reference run of the code to be compared to a current run. PCAST can be used in two modes: automatically running the CPU and GPU version of a kernel then comparing their outputs directly immediately after, or comparing the output to a previously generated PCAST output file. 
%{\color{purple} RDL - Added PCAST at the end of this sentence to distinguish from your own manually generated output file. I think the next sentence needs clarification but I can't fix it: it sounds like applying PCAST globally didn't work but we found a workaround? } \textcolor{blue}{Eric - tried to change the wording a little bit, hope it's more clear now} 
%\textcolor{red}{SC: Rich, updated PCAST text}
PCAST also has the option to generate all of these comparisons automatically with a compiler flag, however given MURaM's code complexities, it was not straightforward to use PCAST.  
%it simply did not work with MURaM, which we account to several complexities within the MURaM GPU code that caused it to fail. 
Regardless of the issues, we were still able to manually apply PCAST throughout the code.

%\vspace{-2em}

To implement PCAST to the code we created a wrapper macro that could be ignored based on a toggle in the code compilation, or if using a non-PGI compiler. This macro was placed before each kernel and captured the data of all of the significant input variables, as well as after each kernel and captured all significant output variables. Then the code was compiled for CPU and ran for only two timesteps to avoid creating too large of a reference file. Any future GPU runs could then be build with PCAST and compared to the CPU reference pointing out any minor discrepancy between the reference run. Additionally, PCAST includes a feature called patching that will replace any incorrect values with their reference, allowing us to see isolated errors and avoid error propagation to later tests.

In the incremental refactoring process, we initially target a single loop nest and apply OpenACC directives to have that loop run on the GPU, without focusing on any sort of optimization. Data management is also handled as locally as possible with all input variables copied to the GPU immediately before the loop and all output variables copied to the host immediately after the loop. This allows for a single isolated portion of the code to be parallelized without having any cascading effects on the rest of the code. The key benefit to this strategy is that we can ensure code correctness first and foremost before moving on to any optimizations, which can introduce new problems into the code. Once that portion of the code has been parallelized, we checked for two things. 1) ensure that the code still produces correct results, 2) verify with a profiler that the portion of code is running on the GPU and is behaving as expected.

Frequent code profiling during development gives a very important sanity check at every step of the process. It ensures that the most recent code changes are running properly on the GPU and contain any expected data movement associated with them. Some problems that this profiling can identify are kernels running significantly slower than expected, extra or unexpected data movement and low kernel performance from metrics such as occupancy or bandwidth-bound kernels. Many of these details become very important once we move past the initial parallelization and move on to optimizing the GPU code.

\subsection{Optimizing Host-Device Data Movement}
After every important loop was running on the GPU we began the process of eliminating redundant data movement between the host (CPU) and device (GPU). These copies are expensive, but an artifact of the careful, incremental nature of our porting strategy. As with any optimization, removing redundant data copies between neighboring kernels required re-verifying code correctness with our validation suite.

We used a GPU profiler, nvprof,  to identify  the data transfers occurring between kernels. Even after we had removed all the unnecessary data transfers from the code we were still observing many small data transfers happening before kernel launches. With the help of the profiler we discovered that device memory allocated with an OpenACC API function works differently than device memory allocated with an OpenACC directive, and causes there to be a small amount of data transfer before kernels where this memory is used. We are not sure exactly what within the PGI compiler causes this, but we now solely use OpenACC directives to eliminate these small data transfers.

Every core MURaM routine 
%used within MURaM for our current configuration 
is ported to the GPU with optimized data movement. Additionally, many computational kernels have been further optimized for GPU performance, and finishing optimizations on the remainder of the code is a clear future direction of the MURaM project.

\subsection{Optimizing GPU-aware MPI}
\label{sec:gpu_mpi}
 
 %\begin{figure}[ht!]
 %       \lstinputlisting[frame=single,language=c,style=myCStyle]{Figures/gpu_aware_MPI.c}
%        \caption{GPU-aware MPI with OpenACC.}
%        \vspace{-1.5em}
%          \label{fig:gpuawarempi}
%\end{figure}

For the multi-GPU runs in this project we are using OpenMPI 3.1.4 that provides support for GPU-to-GPU MPI data transfers when installed on a machine with compatible hardware. To use this feature a valid GPU address pointer is passed into the MPI function call. In a language such as CUDA this is very straightfoward, as the programmer explicitly manages GPU memory allocations. In OpenACC however the GPU memory allocation is hidden from the programmer by the OpenACC runtime. To expose the GPU address pointer OpenACC provides the \textbf{host\_data} directive and \textbf{use\_device} clause that allows interoperability with GPU-based libraries. While using GPU-aware MPI with OpenACC is very simple, verifying that device-to-device data transfers are working as expected is a challenge. The only way that we currently know to verify this functionality is by profiling the MPI GPU application using nvprof and checking explicitly that device-to-device (or DtoD) data transfers are occurring. Within our development system there was a significant amount of trouble getting OpenMPI installed with proper GPU-aware support, and several iterations of testing were required to reach the expected hardware performance level.

\section{Optimizing Radiation Transport}
\label{sec:RTS}
%{\color{purple} RDL - Fact check my terminology regarding multi-spectral vs multi-band, radiative transfer vs radiative transport, etc. } 
It is well known that computing 3D radiation transport (3D-RT) is extraordinarily expensive, depending on two angular dimensions (i.e. the zenith and azimuthal angles) and three spatial dimensions. RT solvers are typically iterative, further adding to the cost. However 3D-RT is absolutely essential in many MHD and astrophysical situations, including solar physics.

% Eric - I moved this clarification to the Implementation section, at the end of when we talk about data optimizations.
%While we dive deeper into only the RTS port in this paper, we also want to add a note that this work does move all the other routines of MURaM also to GPUs. We have not yet fully optimized all the other routines. 

%MURaM uses an explicit RT solver method based on the method of discrete ordinates, which discretizes the angular dimensions. The method is iterative, further adding to its cost, with a convergence rate that depends on the optical depth of the medium, making the overall cost of 3D-RT somewhat dependent on the size of the MPI subdomain.

In the case of MURaM's $288^3$ reference test case with one band, the radiation transport solver (RTS) is the most expensive part of the calculation, accounting for nearly half the time on a single, dual-socket CPU node.
%{\color{purple} RDL - in fact I make it out from the 1 node final weak scaling benchmark on Cobra Skylakes to be 47.7 percent.} 
More realistic, multi-node 3D radiation transport, in which the RTS must be called once per frequency band, will drive the proportion of time spent in RTS even higher.   

For this reason RTS has received the most optimization effort during our GPU port. However, RTS exhibits a wavefront data dependency pattern in its primary computational kernel as well as several blocking MPI communications. This section will focus on optimizing RTS while considering the complexities and limitations of the solver's underlying algorithm.

\subsection{Summary of the Radiation Transport Solver}
 The MURaM code uses short characteristics to solve for the radiation field \cite{Kunasz:Auer:1988}. This method involves integrating the radiation transfer equation along a ray using values of intensity and opacity interpolated from the neighbouring grid cells. To calculate the mean intensity and radiative fluxes at each grid point a set of 3 rays per octant are used to integrate over the unit sphere. The solver uses 3D domain decomposition and iterates the intensities at the boundaries of each sub-domain until the errors are within a prescribed tolerance. A multi-band opacity scheme is used in order to efficiently include the frequency dependence of the radiation transfer problem in the solar atmosphere. For feedback into the MHD code the radiative heating or cooling at each point in space is calculated.
  In practice, using 3 rays per octant results in 24 total rays that are iteratively computed until they converge to an error threshold. 
 %{\color{purple} RDL -we say twice that some routines are run many times - I cut the first reference to that fact}
 %For this reason it is expected that the core computational routines in RTS will be run ~50-100 times for each timestep of the MURaM simulation, whereas most other routines within MURaM only run once per timestep. 
 Figure ~\ref{fig:gpuutilization} shows the structure of the core computational loop in RTS. Expanding on the routines described in Section ~\ref{sec:routines}, \textit{writebuf()} and \textit{readbuf()} pack and unpack a buffer that will be exchanged with neighbors in the \textit{exchange()} routine.
 
 \begin{figure}[ht!]
\begin{lstlisting}[frame=single,language=c,style=myCStyle]
for( < x direction > )
 for( < y direction > )
 for( < z direction > )
 for( < angle > ) // Total 24 rays
 {
    interpol();
    while(g_maxerr > THRESHOLD)
    {
        readbuf();
        integrate();
        writebuf();
        exchange();
        double err = error();
        MPI_Allreduce(&err, &g_maxerr);
    }
    readbuf();
    flux();
 }
\end{lstlisting}
        \caption{Pseudocode of RTS's  core computational loop}
        \vspace{-1.5em}
          \label{fig:gpuutilization}
\end{figure}

\begin{figure}[ht!]
    \centering
    \includegraphics[width=0.3\textwidth]{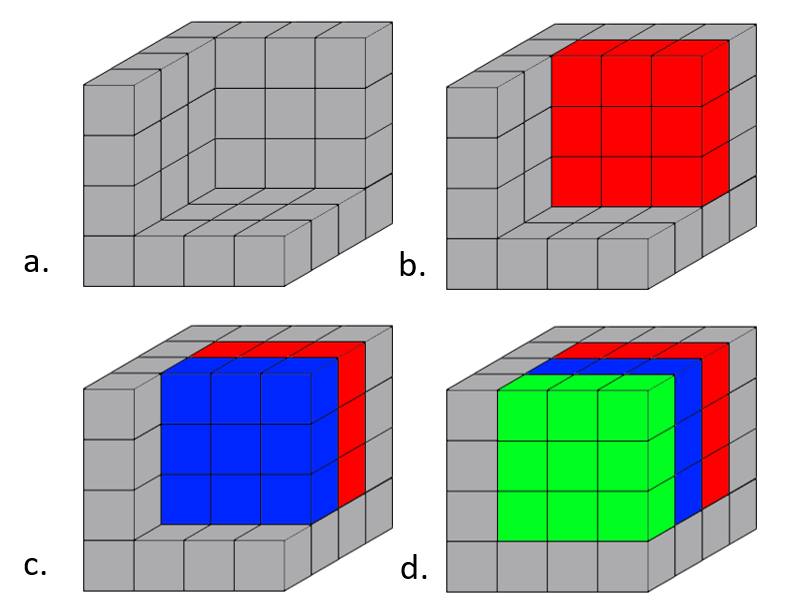}
    \caption{Representation of the wavefront dependency in RTS integration. a) shows the preloaded boundary conditions with no computation done. b) shows the first set of grid points in red to be computed in parallel. c) shows the next set of grid points to be computed in parallel, each blue grid point depends on up to four of the red grid points. d) shows the last grid points to be computed in parallel, each green grid point depends on up to four of the blue grid points.}
    \vspace{-2.1em}
    \label{fig:integrate_dependency}
\end{figure}

 The most important routine within RTS is called \textit{integrate()}, which is executed ~50-100 times per timestep. However, unlike the other routines, \textit{integrate()} introduces a wavefront data dependency pattern that restricts the parallelism attainable within it. This means that even though we are working with a 3-dimensional domain, we can only achieve two dimensions of parallelism; one dimension of our problem will have to remain sequential as displayed in Figure ~\ref{fig:integrate_dependency}.

 %\vspace{-2.1em}
 This data dependency introduces several performance challenges that must be addressed: 1) GPU kernels with 2-D data parallelism are very small and under-utilize the compute processors of GPUs, 2) having one dimension of the domain remain sequential means we must launch possibly hundreds of small GPU kernels to compute the entire domain, introducing runtime-dominating kernel launch overhead, 3) since each ray sweeps in a different direction, some rays have better memory access striding than others, meaning that some rays take several times longer to compute than others.

\label{sec:rts_summary}

\subsection{Reducing Kernel Launch Overhead }
%First, we will explain our methodology for reducing kernel launch overhead. 
Anytime a GPU kernel is launched a certain amount of overhead is incurred. When this overhead is very small compared to the time spent in the GPUs parallel computation, good performance is still achieved. However, in edge-cases such as our \textit{integrate()} routine, this overhead can become a dominant factor in the overall execution time. The OpenACC \textbf{async} directive can be used to hide a large portion of this overhead by queuing many kernels on the GPU before they are run. While the earlier kernels are being computed, the later kernels are being pre-loaded which allows overlap between the execution of the current kernel and the setup of the next one. However, by closely analyzing the GPU profiler, nvprof, it was clear that even with this optimization our \textit{integrate()} routine was still performing sub-optimally due to kernel launch overhead.
 
\begin{figure}[h]
    \centering
    \includegraphics[width=0.44\textwidth]{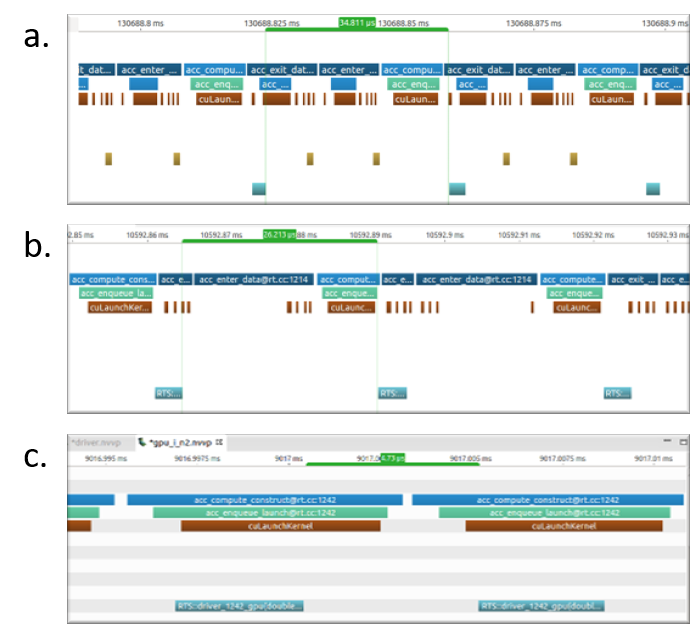}
    \caption{Nvprof profiler highlights the process of reducing kernel launch overhead in the RTS \textit{integrate()} kernel. (a) Overhead between 2 launches, (b) Reduced overhead between kernels launches, and (c) Further reduced overhead}
    \vspace{-1.5em}
    \label{fig:integrate_overhead}
 \end{figure}

 Figure ~\ref{fig:integrate_overhead}a shows the overhead between two launches of the \textit{integrate()} kernel. Each of these kernels are only computing two dimensions of the dataset, and a single call to the \textit{integrate()} routine will result in a few hundred of these kernels being launched. In between the computation of each kernel is ~35µs where the GPU processors are idle. The profiler revealed that there is some amount of data movement happening before and after each kernel, and the source was a pointer translation. This is likely related to the host address pointer needing to be translated with the device address pointer within the OpenACC runtime.
 
 To address this, we changed how the arrays in \textit{integrate()} were being allocated; instead of allocating them on the host first then the device, we allocated them only on the device. The effect of this is seen in Figure ~\ref{fig:integrate_overhead}b, where the overhead has been reduced to ~26µs between kernels. This is an improvement, but the profiler still clearly shows that something is still happening between these kernel launches. We found that this has something to do with how C++ class members are handled. Since all of the arrays used within RTS are stored within a class we created a pointer outside of a class that referenced the needed arrays, and used those new pointers in \textit{integrate()}. After this change Figure ~\ref{fig:integrate_overhead}c shows that the overhead is further reduced to ~5µs between the kernels. This means that \textit{integrate()} is still spending nearly half of its time with idle GPU processors. In order to address this issue, we would need to refactor the code as discussed in the next section.
 
 %but it is not clear that we can achieve better without some code refactoring.
 
\subsection{Restructuring of RTS Integration}
\label{sec:rts_variations}
% Eric - rewording this
% This section will discuss further RTS code optimizations, namely data transposition and integration of interpolation with integration.  

This section will discuss further RTS code optimizations that require refactoring the RTS code.
 
\subsubsection{\textbf{RTS Transpose}}
\label{sec:rts_transpose}
 Since \textit{integrate()} has to deal with a data dependency there are three possible scenarios that we refer to as an x, y or z dependency. These names depend on which of the three dimensions must remain sequential, as outlined in Figure ~\ref{fig:integrate_dependency}. An x or y dependency are parallelized in such a way that the z direction is handled by the inner most loop, which allows for vectorization on a perfectly strided array. However, for the z dependency, the x direction is the inner most loop which results in very bad memory striding. The z dependency is up to 10x slower than x or y, depending slightly on the dataset. The domain decomposition and dataset dimensions will determine how often the rays exhibit the z dependency, and in typical runs it happens often.
 
 Our solution to this was to create a second version of all of the arrays used within \textit{integrate()} to have an alternate transposed version. This includes the intensity which is calculated within \textit{integrate()} and the coefficients which is calculated within \textit{interpol()}. \textit{Interpol()} uses three other arrays to calculate the coefficients, but they are read-only within \textit{interpol()}. We transpose the three read-only arrays in \textit{interpol()} once per timestep before the main computation happens, then we ensure that the coefficients are written in the same order that \textit{integrate()} will read them. Then we transpose the intensity array back before the flux values are calculated.
 
 This results in the \textit{integrate()} kernel being able to work in perfect stride even when encountering a z dependency pattern. The cost is the few extra transposes that have to be done, but in our test runs we are still seeing a significant improvement with \textit{integrate()} running ~5x faster overall, and RTS as a whole running ~3x faster. This is ultimately a performance loss for the CPU, but is easily manageable using branching compilation with ifdef in C++. This "data tranposition" version of RTS performed well for the GPU in every test case we have tried, so the results discussed in Section ~\ref{sec:results} use these changes.
 
 %% Moving onto the merge alternate version
 \subsubsection{\textbf{RTS Integration and Interpolation Merge}}
 \label{sec:rts_merge}
 
 %One strategy for addressing the ~50\% idle time of integrate() is to increase the amount of computations performed in each kernel. 
 %Since the amount of parallelism available in integrate() is limited by a data dependency the kernel has very little work to do and results in low occupancy of the GPU processors. Additionally, since the kernels are so small the processors are idle ~50\% of the time during integrate() as the time-to-compute is shorter than the kernel launch overhead. One possible solution to this is to merge the work of multiple kernels into one.

 One strategy for addressing the ~50\% idle time of \textit{integrate()} is to increase the amount of computations performed in each kernel.
 The two kernels that can be most easily merged are \textit{interpol()} and \textit{integrate()}. However, as seen in Figure ~\ref{fig:gpuutilization} \textit{integrate()} is inside of a convergence while loop and interpol() is outside of it. This is because the coefficients calculated in \textit{interpol()} only need to be done once for each ray and the results are stored within an array to be read from in \textit{integrate()}. Instead of storing the coefficients into an array we moved their calculation into the \textit{integrate()} kernel and used them directly. This means that any ray that takes more than one iteration to converge will now have to compute these coefficients more often, but with the possible benefit that now with more work to do the \textit{integrate()} kernel may be more efficient.
 
 The results of this change varied in our test cases. For smaller datasets, such as $64^3$, where the GPU occupancy was at its lowest and the kernel had less work to do, the \textit{interpol()} and \textit{integrate()} merge resulted a speedup of around 30\% for RTS overall when compared to a GPU run without this change. This optimization might be beneficial for future MURaM forecasting scenarios, where high throughput (capability) would be of paramount importance, requiring strong scaling to smaller per-GPU problem sizes to achieve. However, problem sizes more representative of the current use of MURaM simulations showed a significant performance decrease of up to 50\%.
 
 %\textcolor{magenta}{MR: I can't think of a production setup that small, but I guess we could reach that size easily if we strong scale a larger setup. Would this allow us to push the strong scaling a little bit further? Can we have both options and select them through a pre-processor directive or can it be decided at runtime with a problem size threshold?}

% \textcolor{blue}{Damien: Not on a smaller dataset, but here is why i want to move towards a J-convergence (or in the case of NLTE Source-function convergence), dependency overlap, interpolate + integrate merged scheme. I intend to try to do this next year for "fun".}
 
 In the chromosphere the radiation field can be strongly scattering, meaning the radiation source function is strongly dependent on the intensity. For strong-scattering problems, the radiation transfer problem becomes more non-local, increasing iterations to convergence. In this case it is preferable to use a Gauss-Seidel convergence scheme~ \cite{trujillo1995novel}. As the intensity is integrated along the ray, this scheme will require updating the source function for each point along the ray, and then using the 'new' source function to integrate the intensity at the next point along the ray. This algorithm requires a combined treatment of interpolation and integration. This modification will therefore be of interest to apply the GPU short-characteristics scheme to broader range of stellar problems. For this reason, this variation of RTS may be an important direction for future work.
 %At the moment we do not believe that this variation of RTS is the direction to pursue unless needing to run on an overall small dataset.
 
 % Ray merge
\subsubsection{\textbf{Similar Dependency Overlap}}
 Another challenge that we addressed was the low GPU occupancy observed in \textit{integrate()}. One technique tried was to combine the computation of several rays into a single kernel. Our first approach was to use OpenACC asynchronous programming to queue the work of all 24 rays simultaneously on the GPU. The hope was that the GPU would be able to overlap the computation of multiple rays since a single ray is only using about 10\% of GPU resources.
 
 In practice, we observed through the GPU profiler, nvprof, that this method did allow for a little bit of overlap between the computation of multiple rays, but far less than what we would assume to be theoretically possible. In our experimentation, this method only improved performance by ~5\%. Typically when OpenACC asynchronous programming, GPU computation overlaps with either CPU computation or CPU/GPU data transfers. Additionally, there is no way to lock specific streaming multiprocessors to specific kernels within OpenACC. If this existed it could possibly be a viable solution to the problem we are facing.
 
% \textcolor{blue}{Damien - Eric, Can you speculate why?} \textcolor{purple}{Eric - added a little bit, could expand further. For exapmle: in CUDA you can set a max number of SMs per kernel, which could potentionally help enable the kind of behavior we are hoping for with this async approach. It is not very common, and ultimately against what is considered as CUDA best practices, but is technically possible. OpenACC async isn't usually used to overlap overlap GPU computation like this, but from the profiling I did it is clearly possible, but the overlap is nowhere near as good as I would think is theoretically possible.}
 
% \textcolor{blue}{Damien - This was never completely implemented right? Should it be in here?}
 
 %\textcolor{green}{Eric - It was somewhat implemented. I finished some work on implementing the multi-ray for the computational kernels, and measured the performance difference that it would have on the computational aspect of rt, but never finished the MPI implementation. I wanted to mention it and mention it as a future direction that the code would go.}
 
We have experimented with a few other optimization variations. First, instead of relying on the OpenACC~\textbf{async} clause to overlap computation, rays that exhibit a similar dependency pattern are combined with an outer parallelizable loop. We can identify 6 groups with 4 rays each that will exhibit a similar dependency pattern with the group. This means that the arrays used within the affected RTS computation routines must be increased in size by a factor of 4. This would increase the work done per \textit{integrate()} kernel by a factor of 4 as well. When only considering the computational benefits of this change we observed a reasonable performance benefit. And similar to the methodology in Section ~\ref{sec:rts_merge} this RTS variation received a larger performance increase for smaller datasets, likely when the idle time between kernels and GPU occupancy is at its lowest.

Since many rays would now be overlapped, instead of computed one-at-a-time, the way to determine convergence is changed to a global convergence instead of a per-ray convergence. This could have the added benefit of reducing the total number of iterations needed overall. Additionally, since 4 rays can be computed simultaneously, there will be 4 times fewer \textit{exchange()} function calls and a factor of 24 times less \textit{error()} function calls and associated MPI all-reduce routine calls which could greatly reduce the MPI communication overhead within RTS. Lastly, since half of the rays are moving upward and half moving downward it is possible that we could use 3 groups of 8 rays instead of 6 groups of 4, as rays moving upward vs. downward could still exhibit the same dependency pattern.

% \textcolor{blue}{The global convergence criterion (angle averaged intensity J), is used to calculate the radiative cooling/heating in the atmosphere, which is fed back into the MHD energy equation. This criterion ensures that the important physical quantity is sufficiently accurate, while reducing unnecessary iterations of highly inclined rays that do not contribute strongly to J and are more difficult to converge. The additional cost of using this is keeping the source function and opacity of all rays in memory, or additional interpolation loops in the RT solver() routine.}
 
%\textcolor{purple}{Eric - not sure yet how to incorperate this text in with the narrative, will take another stab later.SC: We can leave it out for this version and put it back into the journal}

Implementing and evaluating these optimization ideas fully remains as future work for the project, and are not utilized when discussing performance results in this paper.

\section{Results}
\label{sec:results}
In this section, we present the results of running MURaM on multiple CPUs and GPUs while demonstrating parallel efficiency. %Results show performance of a single GPU performance, strong scaling, and weak scaling using the Cobra cluster.
\vspace{-0.5em}
\subsection{\textbf{Experimental Setup}}
The Cobra system consists of 3424 compute nodes, each containing two Intel Xeon Gold 6148 Skylake (SKL) processors (20 cores at 2.4 GHz) and 100 Gb/s OmniPath interconnect. There are 64 GPU nodes with 2 NVIDIA Tesla V100-PCIE-32GB per node utilizing 32 GB HBM2 for a total of 7.9 TB HBM2 across all nodes.
%and achieving theoretical peak double precision performance of 11.4 PFlop/s peak double precision along with 2.64 PFlop/s peak single precision. 
CPU runs: Intel 19.1.3, Intel MPI 2019.9, MKL 2020.2, FFTW-MPI 3.3.8. 
GPU runs: For the -O0 runs - NVHPC 20.4, CUDA 10.2, OpenMPI 4.0.5 with UCX 1.8.0, FFTW-MPI 3.3.8; For the -O3 runs - NVHPC 20.9, CUDA 11, OpenMPI 4.0.5 with UCX 1.8.0 and FFTW-MPI 3.3.8. 
At the time of submission,we were able to collect results for GPU strong scaling  for the $288^3$ dataset using -O3. For the weak scaling, the runs used -O0 and the -03 runs were finishing up. Preliminary results showed for the $288^3$ dataset, with -03, the GPU weak scaling ran approximately 10\% faster. 
(Note: PGI compilers have been since October 2020 renamed to NVHPC~\cite{nvhpc})
%\textcolor{purple}{Eric - we mention that some runs are -O3, some are -O0, should we mention why we could not get all -O0 in time?SC: Added a kinda sorta blurb}
%\vspace{-1.5em}
\subsection{Single GPU Performance}
 Our initial project focus was optimizing performance on a single GPU. In practice, the simulations that are run with MURaM will require multiple GPUs, but using a smaller dataset allows us to study GPU performance separately from the scalability of the MPI implementation. Using a single V100 GPU, the average time to simulate one single-band (or gray scale) timestep with the $288^3$ dataset is 2.285 seconds. This is a 1.73x speedup over the 3.949 seconds required to run the same simulation on a fully subscribed 40-core Skylake CPU node. 

\subsection{Strong Scaling}
\begin{figure}[t]
\centering
    \includegraphics[width=0.9\columnwidth]{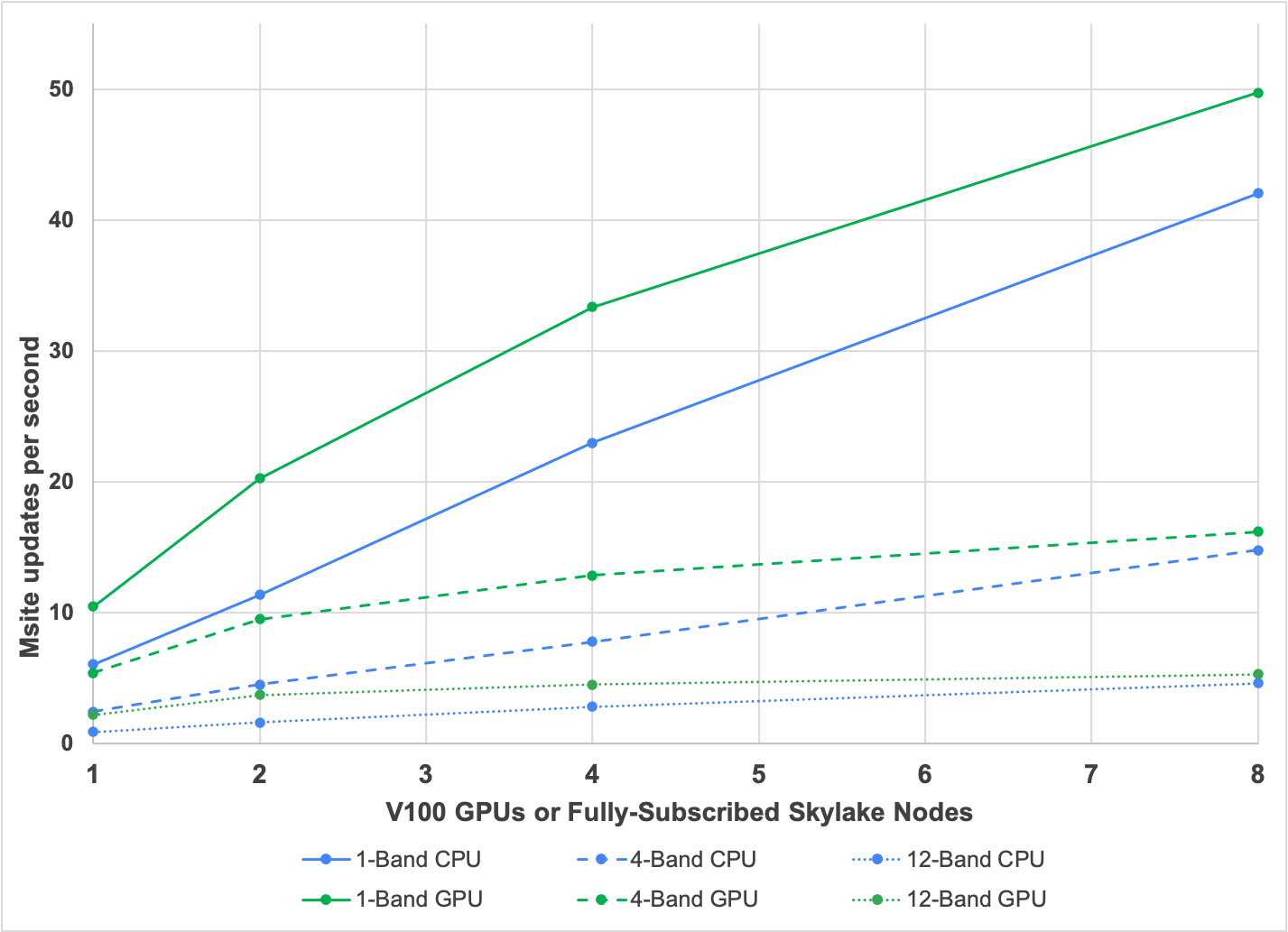}
    \caption{CPU-GPU strong scaling of a multi-band $288^3$ dataset}
    \vspace{-1.5em}
    \label{fig:StrongScalingCobra}
\end{figure}

%\textcolor{teal}{CEM: Updating with the 12 bands, and changing the language! SC: Brilliant, thanks Cena. Removed my comment}
Strong scaling is defined as how the solution time varies with the number of processors for a fixed total problem size. For our strong scaling experiment we use a $288^3$ dataset divided across 8 GPUs and 8 fully subcribed CPU Skylake nodes (40-cores per node). 
The CPU runs use -O3 and AVX512 flags and the GPU runs use -O3. %\textcolor{teal}{CEM: FYI, We only have -O3 GPU strong scaling results, and -O0 GPU weak scaling results as of Sun morn. }
We are also comparing the strong scaling performance of the code with gray band (1 band) and colored band (4 band and 12 band). The colored band increases the workload in RTS and is proportional to the scaling of RTS. 
Figure ~\ref{fig:StrongScalingCobra} shows the strong scaling of MURaM with this configuration. 
The performance of the code is measured in millions of sites updates per second (Msite/s). %\textcolor{red}{SC: We need to define what is Msite updates per second}
%Throughput is defined as seconds simulated time per second of execution, and we draw a comparison of 1 GPU vs. a fully subscribed CPU node.
 Cobra contains 2 GPUs per node, so increasing to 4 GPUs requires inter-node communication. From 1 to 2 GPUs the code scales very well in both the single and 4-band case, roughly doubling the throughput. However, moving from 2 to 4 GPUs only increases throughput by 1.63, which is possibly due to the higher cost of internode communication. Preliminary results shown for the 8 GPU runs indicate that a more optimal core decomposition for multi-node scaling may be chosen after further testing. 
 
\begin{figure}[t]
\centering
    \includegraphics[width=0.9\columnwidth]{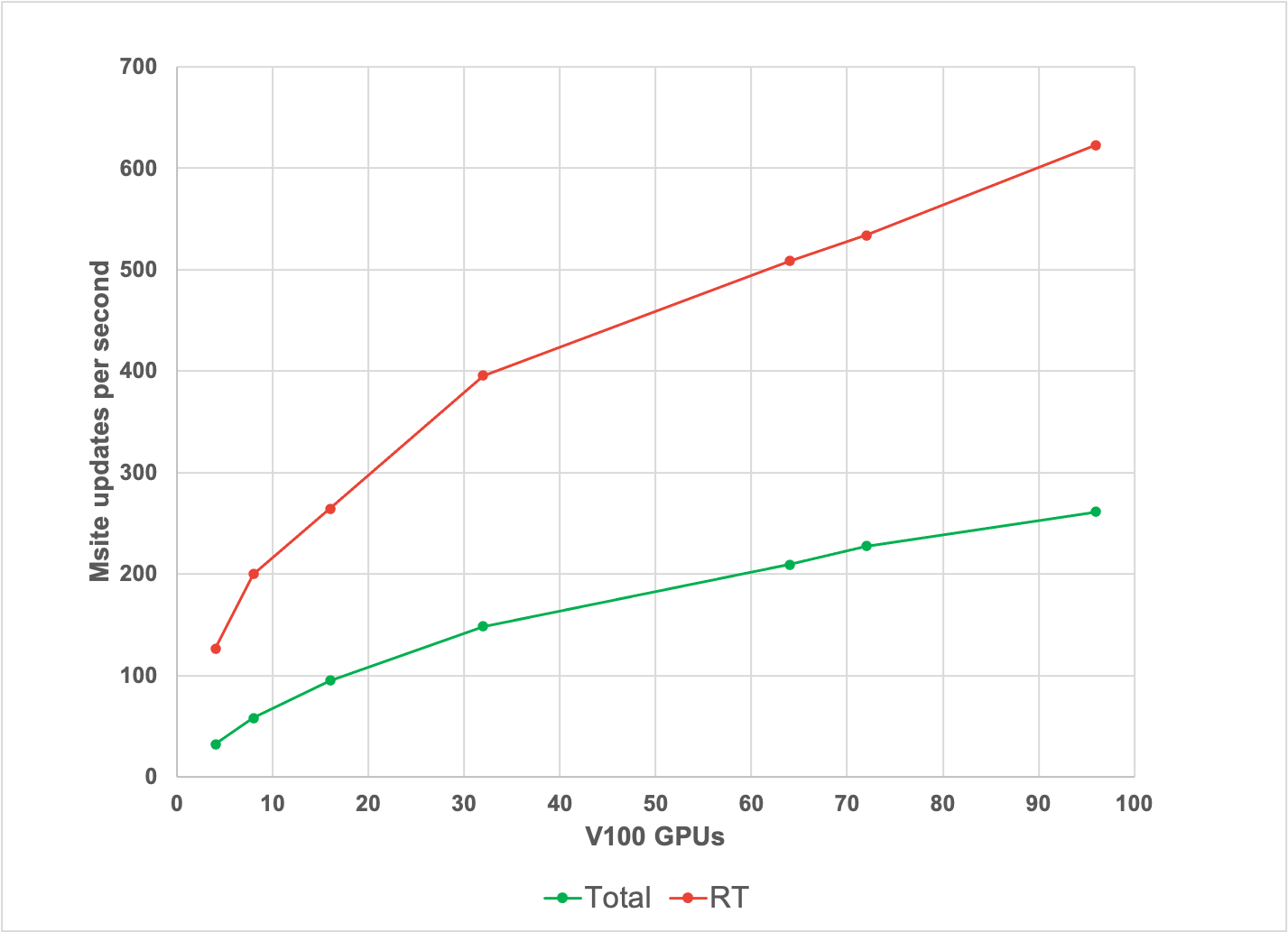}
    \caption{GPU Strong Scaling of a $288x576x576$ dataset}
    \vspace{-1.5em}
    \label{fig:StrongScalingCobra576RT}
\end{figure}

 We also gathered strong scaling results using up to 96 GPUs for a $288*576*576$ single-band dataset, as shown in Figure ~\ref{fig:StrongScalingCobra576RT}. These results show the strong scaling performance of the RTS routine as compared to the overall simulation: both seem to do reasonably well. The scaling efficiency of both RTS and the full model begin to decrease when the dataset is split over more than 32 GPUs, and the number of points per GPU falls below one million. 
%\textcolor{red}{SC: CEna here... we talk about the new Fig.12}

%\textcolor{purple}{Eric - I think this is the first time we have mentioned bands. Don't know if that's a problem, just pointing it out.}

\subsection{Weak Scaling}
 %For our weak scaling results we allocate a $288^3$ dataset to each GPU/fully subscribed CPU node. The general use case of MURaM often involves problems that scale horizontally rather than vertically, so in our weak scaling we are expanding the horizontal directions of the problem rather than the vertical direction. Figure ~\ref{fig:WeakScalingCobra} shows the weak scaling comparison of up to 100 GPUs (50 nodes) vs. 100 fully subscribed CPU nodes (4,000 cores). The graph is measured in seconds per timestep which is the average of the first 100 iterations of the simulation, excluding I/O and initialization. Overall, RTS does not scale particularly well in the CPU code which results in the time taken gradually increasing when adding more nodes, with 100 CPU nodes being 16\% slower. The GPU code scales significantly better, with 100 GPUs performing only 7\% slower than 64 GPUs.
%\textcolor{purple}{Eric - putting it into percents like this makes it look less impressive than the graph visual. Not sure what the best way to describe the information is.SC: Agreed. Let us put them in numbers in seconds and see how it reads.}

\begin{figure}[t]
\centering
    \includegraphics[width=0.9\columnwidth]{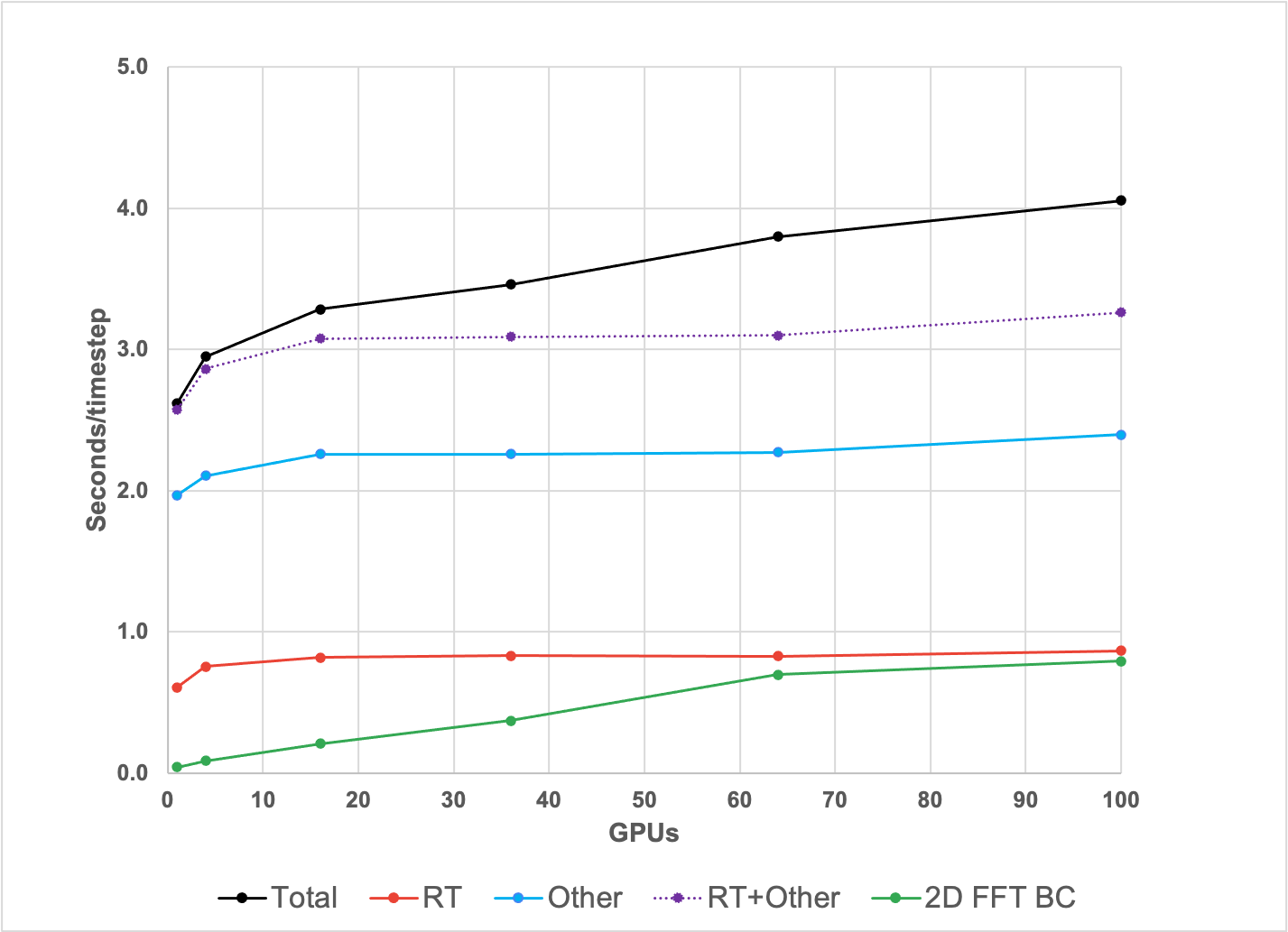}
    \caption{GPU weak scaling of a $288^3$ dataset for RTS and other MURaM routines}
    \vspace{-1.5em}
    \label{fig:WeakScalingCobra_breakdown}
\end{figure}

Weak scaling is defined as how the solution time varies with the number of processors for a fixed problem size per processor and gives a great deal of information about the MPI communication and overall scalability of the code. Figure ~\ref{fig:WeakScalingCobra_breakdown} shows a breakdown of the GPU weak scaling with respect to the different routines of the code. This figure is also measured in terms of seconds per timestep. We use a $288^3$ dataset per GPU for this experiment. It is clear that between 1 to 64 GPUs the scalability suffers due to the FFTW library. Currently, we are using multi-threaded FFTW  on the CPU with results being copied to the GPU after computations. As a future direction, we plan to explore a GPU-enabled FFT library.

\subsection{Results Summary}
%\textcolor{red}{SC: LOVE this text! Thank you to whoever wrote this.}
  There is an increased computational cost moving the simulation from single-band to multi-band, as seen in Figure ~\ref{fig:StrongScalingCobra}. For N-bands, the interpolation, integration, and flux calculation routines within RTS run N times more often. 
  %Additionally, there is a difference between single- and multi-band in the averaging method. 
  Additionally, there are differences in the convergence rate of RTS. Single-band RT is more transparent in the highly structured upper photosphere, resulting in good scaling in Figure~\ref{fig:StrongScalingCobra}. Multi-band RT is more sensitive to those structures and converges more slowly.
  %For a single-band, all wavelengths in the solar spectrum are averaged together, resulting in relatively smooth opacities and we see good scaling in Figure~\ref{fig:StrongScalingCobra}. For a multi-band, the spectrum is split based on wavelength strength in the atmosphere. This results in different regions of the atmosphere with more widely varying opacity, increasing the non-locality and therefore the iterations to convergence; a likely reason why the multi-band, for example the 4-band does not scale well in the same figure. 

\section{Related Work}
Accelerating the solar physics simulation models through GPU devices is a relatively new direction which attracts broad interest from the scientific community. 
%Recent studies include researchers~\cite{Caplan1} presenting an example of accomplishing the production level solar physics simulation on a single multi-GPU server at speeds used to requiring large batch jobs on supercomputer. 
Related work~\cite{caplan2019gpu} shows the "Time-to-solution" performance results of a flux rope eruption simulation with their OpenACC implementation of MAS code. MAS represents Magnetohydrodynamic Algorithm outside a Sphere(MAS), an in-production MHD code, which is part of the CORHEL suite hosted at the Community Coordinated Modeling Center (CCMC). Note that MHD is also a kernel in MURaM. The same group of researchers also summarized the implementation of OpenACC into MAS, including specific code example, strategies and development tips. Other non-GPU MHD codes include BATS-R-US~\cite{powell1999solution} that solves 3D MHD equations in finite volume form using numerical method. The code uses Message Passing Interface (MPI) and the Fortran90 standard. Similary PLUTO~\cite{mignone2007pluto}, a numerical code for computational astrophysics is a multi-physics, multi-algorithm, high resolution framework that comprises of MHD, which is one of the four independent physics modules of PLUTO. 

As presented in this manuscript, one of the major kernels of interest to us in MURaM is the radiation transport. Related work include the recent acceleration of minisweep, a MiniApp of the Denovo~\cite{evans2015three} radiation transport application on GPUs~\citep{searles2019mpi+,searles2018abstractions} using OpenACC. Results demonstrate that OpenACC running on NVIDIA’s next-generation Volta GPU boasted an 85.06x speedup over serial code, which is larger than CUDA’s 83.72x speedup over the same serial implementation. 
Other MiniApps demonstrating radiation transport approaches include Kripke~\citep{kripke} that uses RAJA~\cite{hornung2014raja} and TestSNAP~\cite{thompson2015spectral} (mimicking communication patterns of PARTISN~\citep{partisn} transport code) investigating different data layout patterns and parallelism using Kokkos and OpenMP offload model~\cite{edwards2013kokkos}. 
Coarray Fortran-based Sweep3D's comparable performance to that of the MPI is discussed in~\cite{coarfa2006experiences}. The work on Ardra~\cite{kunen2019porting} discusses porting discrete ordinate transport code to CUDA while using RAJA model with CHAI~\cite{chai} and Umpire~\cite{umpire} to manage multiple memory spaces.
\vspace{-0.5em}
%Different algorithm options are analyzed. The researchers focused on the "zero-beta" mode which freezes the density and sets the pressure to zero, essentially eliminating several evolution equations. They also chose the Runge-Kutta Legendre super time-stepping(STS) scheme and the inverse diagonal of the matrix (point-Jacobi) type preconditioner(PC1). The STS mainly consists of stencil type operations and PC1 is a simple array operation, which both are suitable for OpenACC GPU acceleration. The procedure and code examples of the OpenACC implementation are introduced. The researchers presented the hardware comparisons among 5 CPU choices across 3 CPU generations, and 3 GPU generations (K80, P100, V100), and 14 single system comparisons including SandyBridge, Skylake CPU, and TitanXP, K100, V100 GPU.

\section{Conclusion}
%\textcolor{purple}{Rich - Please vet this pack of \textit{lies and misnomers}, before submitting}

We have successfully added GPU acceleration to the MURaM code using OpenACC using a test-driven refactoring methodology. The refactored code achieves performance-portability between CPU and GPU, and with one exception (the RTS transposition optimization described in Section~\ref{sec:rts_transpose} above), is implemented as a single code-base. While our work represents a fully functional port of MURaM to GPUs, not all of MURaM's routines have been fully optimized. 
Rather, we have focused here on a series of optimizations to RTS,  the most expensive and therefore most critical routine to MURaM performance. 
Results for a $288^3$ test problem show that MURaM with the optimized RTS routine achieves 1.73x speedup using a single MVIDIA V100 GPU over a fully subscribed 40-core Intel Skylake CPU node and with respect to the number of simulation points (in millions) per second, a single NVIDIA V100 GPU is equivalent to 69 Skylake cores.
Multi-GPU weak scaling tests on Cobra show that all components of MURaM, with the exception of the 2D FFT call in BND, can be scaled out. This opens the door to MURaM experiments on large solar domains once a more scalable 2D FFT is identified and incorporated. Strong scaling tests illustrate the point that GPUs can run RTS with 4 frequency bands as fast or faster than comparable numbers of CPU nodes can run single band, enabling routine use of multi-band RTS in solar research.  The OpenACC implementation will allow the MURaM code to utilize the latest state-of-the-art HPC multi-GPU systems deployed at various leadership-class centers around the world.  We fully expect further improvements to the model's performance and scalability as our optimization focus widens to include other routines and as the next generation of GPUs are deployed.

%We plan to continue the optimizing and improving the scalability of  OpenACC implementation of MURaM, and creating new GPU-enabled algorithms needed for improved chromospheric simulations. 

%\textcolor{red}{TO ADD: This paper demonstrates a full port of MURaM. We primarily optimize RTS. We note that there is more room for optimizations of all the other routines in the code.}

%\textcolor{red}{TO ADD: Future Work: The new cluster will consist of 700 40GB A100's with NVlink and Mellanox interconnect. This opens up a number of opportunities to experiment MURaM and explore a number of ways to optimize the code as well.}

%\section{Acknowledgement}
%\input{Ack/ack}

%\clearpage
\bibliographystyle{ACM-Reference-Format}
\bibliography{paper,papref_m}

\end{document}